\documentclass[twocolumn]{article}
\usepackage[utf8]{inputenc}
\usepackage{amsfonts}
\usepackage{amsmath}
\usepackage{amsthm}
\usepackage{amssymb}
\usepackage{mathrsfs}
\usepackage{graphicx}
\usepackage{caption}
\usepackage{ dsfont }
\usepackage{ esint }
\usepackage{color,xcolor}

\theoremstyle{plain}

\theoremstyle{definition}

\newtheorem{remark*}[]{Remark}

\newcommand{\laplacian}{\Delta}

\newcommand{\divergence}{\text{div}}
\newcommand{\gradient}{\nabla}

\title{\bf Stability of Contraction-Driven Cell Motion}
\author{C. A. Safsten$^\dagger$, V. Rybalko$^\ddagger$, L. Berlyand$^\dagger$}
\date{{\small\it $^\dagger$ Department of Mathematics, Pennsylvania State University, University Park, PA	16802}\\
	{\small\it $^\ddagger$ B. Verkin Institute for Low Temperature Physics and Engineering of NASU, 47 Nauky ave, Khariv 61103, Ukraine.}\\
	\vspace{.1in}
	 {\small E-mail: lvb2@psu.edu}}

\begin{document}

\twocolumn[
\begin{@twocolumnfalse}
	\maketitle
	\begin{abstract}
		We consider motility of keratocyte cells driven by myosin contraction and introduce a 2D free boundary model for such motion. This model generalizes a 1D model from \cite{RecPutTru2013} by combining a 2D Keller-Segel model and a Hele-Shaw type boundary condition with the Young-Laplace law resulting in a boundary curvature term which provides a regularizing effect. We show that this model has a family of traveling solutions with constant shape and velocity which bifurcates from a family of radially symmetric stationary states. Our goal is to establish observable steady motion of the cell with constant velocity. Mathematically, this amounts to establishing stability of the traveling solutions. 
		Our key result is an explicit asymptotic formula for the stability-determining eigenvalue of the linearized problem. This formula greatly simplifies the task of numerically computing the sign of this eigenvalue and reveals the physical mechanisms of stability.
		The derivation of this formula is based on a special ansatz for the corresponding eigenvector which exhibits an interesting singular behavior such that it asymptotically (in the small-velocity limit) becomes parallel to another eigenvector. 
		This reflects the non-self-adjoint nature of the linearized problem, a signature of living systems. Finally, our results describe the onset of motion via a transition from unstable radial stationary solutions to stable asymmetric traveling solutions.
		\vspace{.3 in}
	\end{abstract}
\end{@twocolumnfalse}
]

\section{Introduction} 

Sustained motion on a substrate has been observed in experiments on living cells. Keratocytes in particular frequently exhibit motion. They are found naturally moving on flat surfaces, e.g., the human cornea, making them ideal subjects for experiment. Moreover, their flat shape lends itself toward two dimensional modeling. Keratocytes are often observed in a stationary state with a circular shape, or traveling with constant velocity and maintaining a constant, asymmetric shape. This motion is explained by three mechanisms: adhesion, protrusion, and contraction, whose effects are summarized as follows. The moving cell contains actin and myosin proteins. Actin polymerizes, forming a \emph{cytoskeleton} which provides structure for the cell. Actin polymerizing near the edge of the cell causes protrusions of the cell membrane. These protrusions then adhere to the substrate, stabilizing the cell in its new shape. Myosin causes the actin polymers to contract. If the myosin is concentrated on one side of the cell, the cell contracts on that side, driving intracellular fluid to the other side of the cell and expanding the cell on that side. The flow of intracellular fluid also carries the myosin to the other side of the cell, continuing the process and resulting in net motion. The study of cytoskeloton gel has led to the recent development of the so-called ``active gel physics'' \cite{ProJulJoa2015}.

We propose a two dimensional free boundary PDE model for cell motility which describes the evolution of the cell shape and the distribution of myosin within the cell. As in \cite{RecPutTru2013}, our model explains cell motility as being driven primarily by contraction (as opposed to adhesion or protrusion). This model exhibits bifurcation of a family of traveling solutions (modeling cells moving with constant velocity and shape) from a family of stationary solutions (modeling non-moving cells).

Of particular interest is the question of stability of traveling solutions to this model. In order to be observed in experiment, steady cell motion must be robust, not disrupted by small perturbations present in any experimental setup. We show that traveling solutions to our model also have this property by showing that they are mathematically stable. Stability is also important for numerical computations. Since any computational model of a moving cell is necessarily an approximation of a true cell, stability of traveling solutions is necessary for numerical simulations of cell motion to converge.

A 1D contraction-driven free-boundary model is proposed in \cite{RecPutTru2013,RecPutTru2015}, see more recent work \cite{PutRecTru2018}. Our model generalizes this one to two dimensions, and also answers the question of stability of traveling solutions. 
Our analysis shows that asymmetry in the myosin distribution results in the net motion of the cell. 
The essence of this phenomenon lies in the ``motor effect'' studied in \cite{PerSou2009} in the context of a 1D model for myosin moving along filaments.

A 2D free-boundary model for cell motion driven by polymerization of actin (as opposed to myosin contraction) is proposed in \cite{BlaCas2013}. Like our model, this model also possesses a branch of traveling solutions bifurcating from a family of stationary solutions. Analysis shows that the bifurcation in this model is subcritical, meaning that traveling solutions near the bifurcation point are unstable (see also a 2D model in \cite{BerFuhRyb2018}).

Other 2D free boundary models of cell motility have been examined numerically, e.g., \cite{BarLeeAllTheMog2015,BlaCas2013,CalJonJoanPro2008}. For example, in \cite{BarLeeAllTheMog2015}, the authors propose a model for keratocyte motility taking into account actin polymerization in addition to myosin-driven contraction. Numerical analysis of this free boundary model shows close agreement with both experimental results and theoretical results in our model. Additionally, a 2D moving cell model where the boundary has fixed shape was introduced and studied analytically and numerically in \cite{EtcMeuVoi2017}. This model possesses several stationary solutions whose stability is proved provided the total myosin mass is sufficiently small. 

Phase-field models of cell motion provide an alternative to free boundary models. Computational results of these models, shown in, e.g., \cite{ZieAra2015}, also agree qualitatively with results from our free boundary model.


\section{The Model} We consider a 2D model for a cell occupying a region $\Omega(t)$ with free  boundary. We model the flow of the acto-myosin network as a gel obeying Darcy's law $-\nabla p=\zeta u$,
where $-p$ is  scalar stress ($p$ being pressure), $u$ is flow velocity, and $\zeta>0$ is the constant adhesion coefficient. We take for the constitutive equation for scalar stress $-p=\mu\,\divergence u+km-p_h$
where $\mu>0$ is the constant bulk viscosity of the gel, making $\mu\,\divergence u$ the hydrodynamic stress; $m$ is the density of myosin with constant contractility coefficient; and $p_h$ is the constant hydrodynamic pressure at equilibrium. We will assume that $\mu$ and $k$ are scaled such that  $\mu=k=1$. 

Following the Young-Laplace law, we assume that on the boundary
$p+p_\ast=\gamma\kappa$ where $\kappa$ is the curvature of the boundary $\partial\Omega(t)$, $\gamma$ is the constant surface tension coefficient, and $p_e$ is the effective elastic restoring force induced by membrane cortex tension. With the idea of generalizing Hooke's law for 1D springs, $p_e$ models the elastic restoring force nonlocally as proportional to the difference between the area $|\Omega|$ and a reference area $|\Omega_h|$:
\begin{equation}\label{eq:restoring_force}
p_e=-k_e\frac{|\Omega|-|\Omega_h|}{|\Omega_h|},
\end{equation}
where $k_e$ is constant inverse compressibility coefficient. The purpose of $p_e$ is to serve as a regularization term, preventing $\Omega$ from becoming arbitrarily large or collapsing to a point (c.f. vertex models, e.g. \cite{AltGanSal2015}).

The evolution of the density of myosin is described by the advection-diffusion equation $\partial_t m=\laplacian m-\divergence(um)$
with the no-flux boundary condition $\partial_\nu m=0$
where $\nu$ is the outward normal vector to the moving boundary $\partial\Omega(t)$. Finally, the motion of the cell boundary is described a kinematic boundary condition $V_\nu=-\partial_\nu p/\zeta$, where $V_\nu$ is the velocity of the boundary in the normal direction. This boundary condition ensures that the normal velocity of the boundary matches the normal velocity $
\nu\cdot u$ of the intracellular fluid. 

Combining the above equations and introducing the convenient potential $\phi=-(p+p_h)/\zeta$, we arrive at the free-boundary PDE model
\begin{align}
0&=\laplacian\phi+m-\zeta\phi && \text{ in }\Omega(t)\label{eq:phi_PDE_1}\\
\partial_t m&=\laplacian m-\divergence(m\gradient \phi) && \text{ in }\Omega(t)\label{eq:m_PDE_1}\\
\zeta\phi&=p_\ast(|\Omega(t)|)-\gamma\kappa && \text{ on }\partial\Omega(t)\label{eq:phi_BC_1}\\
\partial_\nu m&=0 && \text{ on }\partial\Omega(t)\label{eq:m_BC_1}\\
V_\nu&=\partial_\nu\phi && \text{ on }\partial\Omega(t)\label{eq:kinematic_BC_1},
\end{align}
where $p_\ast=p_h+p_e$.

The system \eqref{eq:phi_PDE_1}-\eqref{eq:kinematic_BC_1} has a family of solutions corresponding to stationary cells. These solutions have constant stress $\phi_0$ and myosin density $m_0$, and a circular shape:
\begin{align}
\phi_0&=\frac{p_\ast(\pi R^2)-\gamma/R}{\zeta}\label{eq:stationary_phi}\\
m_0&=p_\ast(\pi R^2)-\gamma/R\label{eq:stationary_m}\\
\Omega_0(t)&=\{(x,y)\in\mathbb R^2:\sqrt{x^2+y^2}<R\}.\label{eq:stationary_shape}
\end{align}
The family of stationary solutions is parameterized by the radius $R$ of the cell. The total myosin mass of the stationary solution with radius $R$ is
\begin{equation}\label{eq:myosin_radius_relation}
M(R)=\pi R^2m_0=p_\ast(\pi R^2)\pi R^2-\pi \gamma R.
\end{equation}
Provided
\begin{equation}\label{eq:monotinicity_condition}
p_\ast'(\pi R^2)<-(2m_0+\gamma/R)/(2\pi R^2),
\end{equation}
the function $M(R)$ is strictly decreasing, so it has an inverse $R(M)$. Therefore, either $M$ or $R$ may be used to parameterize the family of stationary solutions.


\section{Traveling Solutions and Bifurcation}

A traveling solution $(\phi,m,\Omega)$ to \eqref{eq:phi_PDE_1}-\eqref{eq:kinematic_BC_1} with velocity $\mathbf V\in\mathbb R^2$ has the ansatz $\phi=\phi(\mathbf x-\mathbf Vt)$, $m=m(\mathbf x-\mathbf Vt)$ where $\mathbf x\in\mathbb R^2$, and $\Omega=\Omega_0+\mathbf Vt$. Substituting this ansatz into \eqref{eq:phi_PDE_1}-\eqref{eq:kinematic_BC_1}, we find that \eqref{eq:phi_PDE_1}, \eqref{eq:phi_BC_1}, and \eqref{eq:m_BC_1} are unchanged while \eqref{eq:m_PDE_1} and \eqref{eq:kinematic_BC_1} become
\begin{align}
\partial_t m&=\mathbf V\cdot\nabla m+\laplacian m-\divergence(m\gradient \phi),\text{ and}\label{eq:moving_m_PDE}\\
0&=\partial_\nu(\phi-\mathbf V\cdot \mathbf x)\label{eq:moving_kinematic_BC}
\end{align}
respectively. Solutions to \eqref{eq:moving_m_PDE} have the form $m=\Lambda e^{\phi-\mathbf V\cdot \mathbf x}$ for some $\Lambda$ depending on $\mathbf V$. Substituting this solution for $m$ reduces \eqref{eq:phi_PDE_1}-\eqref{eq:kinematic_BC_1} to the system
\begin{align}
0&=\laplacian\phi+\Lambda e^{\phi-\mathbf V\cdot \mathbf x}-\zeta\phi && \text{ in }\Omega\label{eq:TW_phi_PDE}\\
\zeta\phi&=p_\ast(|\Omega|)-\gamma\kappa && \text{ on }\partial\Omega\label{eq:TW_phi_BC}\\
0&=\partial_\nu(\phi-\mathbf V\cdot\mathbf  x) && \text{ on }\partial\Omega\label{eq:TW_kinematic_BC}
\end{align}
for unknowns $\phi$, $\Omega$, and $\Lambda$, each of which depend on $\mathbf V$. When $\mathbf V=0$, we recover the stationary solution \eqref{eq:stationary_phi}-\eqref{eq:stationary_shape}, where $\Omega$ is a disk of radius $R$. Therefore, for $\mathbf V\neq 0$, we take the cell boundary $\partial\Omega$ as the polar graph of the curve $R+\rho(\theta,\mathbf V)$ where $\rho(\theta,0)=0$. Choose coordinates such that $\theta=0$ corresponds to the direction of $\mathbf V$ so $\phi$, $\rho$, and $\Lambda$ depend only on $V=|\mathbf V|$. To approximate solutions to \eqref{eq:TW_phi_PDE}-\eqref{eq:TW_kinematic_BC} we take asymptotic expansions of the unknowns as follows:
\begin{align}
\begin{split}
\phi(r,\theta&,V)=\phi_0(r,\theta)+\phi_1(r,\theta) V+\phi_2(r,\theta) V^2\\
&+\phi_3(r,\theta) V^3+O(V^4)
\end{split}\label{eq:phi_expansion}\\
\rho(\theta,V)&=\rho_1(\theta)V+\rho_2(\theta)V^2+\rho_3(\theta)V^3+O(V^4)\label{eq:rho_expansion}\\
\Lambda(V)&=\Lambda_0+\Lambda_1 V+\Lambda_2 V^2+\Lambda_3 V^3+O(V^4).\label{eq:Lambda_expansion}
\end{align}

Substituting the expansions \eqref{eq:phi_expansion}-\eqref{eq:Lambda_expansion} into \eqref{eq:TW_phi_PDE}-\eqref{eq:TW_kinematic_BC}, we obtain coefficients $\phi_n$, $\rho_n$ and $\Lambda_n$ via an iterative procedure. As part of this procedure, we take into account specific features of the free boundary, e.g we transform the free boundary condition \eqref{eq:TW_phi_BC} for $\phi(R+\rho)$ to a fixed boundary condition for $\phi(R)$ and its derivatives by expanding $\phi$ about $r=R$. Solving for $\phi_n$, $\rho_n$, and $\Lambda_n$, we find that $\phi_0$ is given by \eqref{eq:stationary_phi}, $\phi_1$ is the product of an explicitly known function of $r$ with $\cos\theta$, and $\rho_1$ is zero. For $n\geq 2$, $\phi_n$ is the sum of Fourier modes up to $\cos(n\theta)$. We solve for the Fourier coefficients numerically.

\begin{figure}
	\centering
	\includegraphics[width=3in]{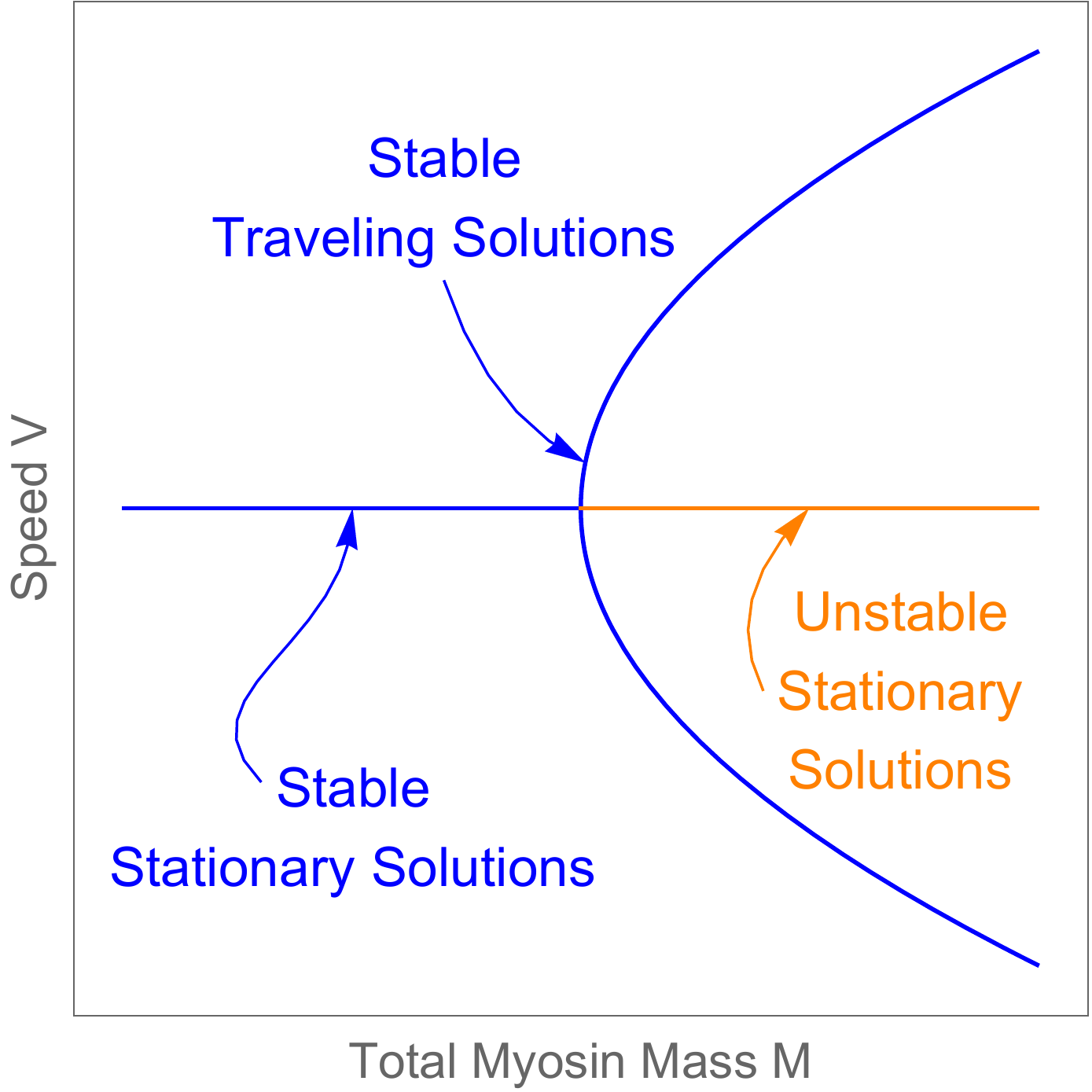}
	\caption{Forward bifurcation diagram ($C>0$) showing the total total myosin mass $M$ and speed $V$ of steadily moving cells. Total myosin mass increases to the right. Blue corresponds to stability, orange to instability.}
	\label{fig:forward_bifurcation}
\end{figure}

\begin{figure}
	\centering
	\includegraphics[width=3in]{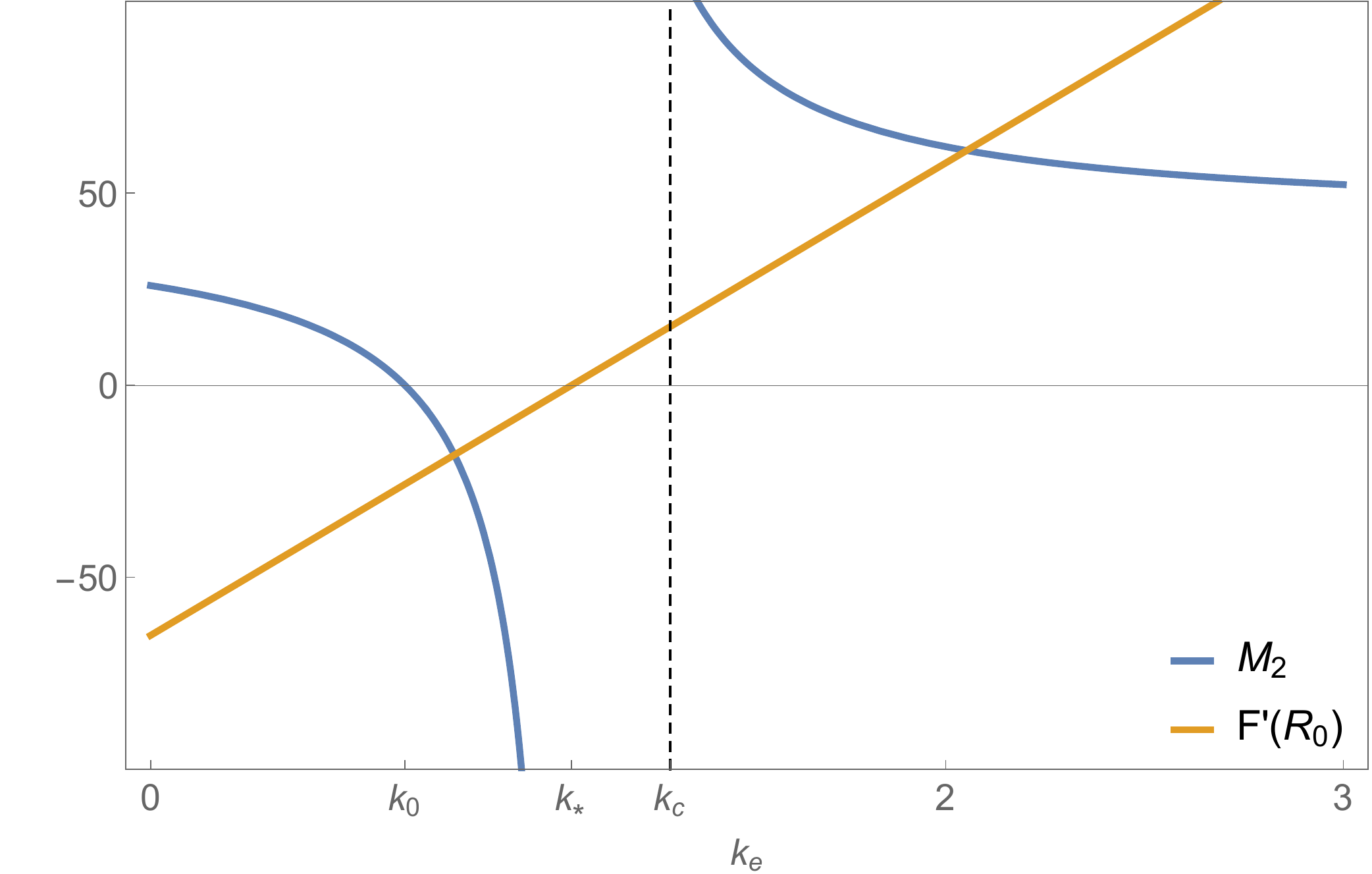}
	\caption{The dependence of the $V^2$ coefficient $C$ of the total total myosin mass $M$ of steadily moving cells with velocity $V$ on the bulk elasticity $k_e$.}
	\label{fig:second_myosin_coefficient}
\end{figure}

The coefficient $\phi_1$ is instrumental for determining at what parameter values the branch of traveling solutions bifurcates from the family of stationary state solutions. All of the coefficients $\phi_n$ are determined by solving PDEs in a disk of radius $R$ with two boundary conditions derived from \eqref{eq:TW_phi_BC} and \eqref{eq:TW_kinematic_BC}. For $n\geq 2$, appropriate choices of $\rho_n$ and $\Lambda_n$ allow $\phi_n$ to satisfy both boundary conditions, but $\rho_1$ and $\Lambda_1$ are decided by other considerations\footnote{The system \eqref{eq:phi_PDE_1}-\eqref{eq:kinematic_BC_1} does not have unique solutions; if $\phi(\mathbf x,t)$, $m(\mathbf x,t)$ and $\Omega(t)$ solve \eqref{eq:phi_PDE_1}-\eqref{eq:kinematic_BC_1}, then so does the translation  $\phi(\mathbf x-\mathbf y,t)$, $m(\mathbf x-\mathbf y,t)$ and $\Omega(t)+\mathbf y$. Therefore, we take $\rho_1=0$ to select the traveling solution centered at the origin. Furthermore, we expect $\Lambda(V)$ to be an even function of $V$: $\Lambda(V)=\Lambda(-V)$, so $\Lambda_1=0$}. Therefore the parameters on which $\phi_1$ depends must be chosen so that $\phi_1$ meets both boundary conditions. Those parameters are $R$ and the physical parameters $\zeta$, $\gamma$, $k_e$ and $p_h$. Specifically, $R$ and the physical parameters must meet the \emph{bifurcation condition}
\begin{equation}\label{eq:bifurcation_condition_1}
F(R,\zeta,\gamma,k_e,p_h):=\frac{\zeta I_1(Rs)}{s^3 I_1'(Rs)}-\frac{m_0 R}{s^2}=0,
\end{equation}
where $s=\sqrt{\zeta-p_\ast(\pi R^2)+\gamma/R}$, and $I_1$ is the modified Bessel function of the first kind with order 1, and $I_1'$ is its derivative. For each choice of the physical parameters, \eqref{eq:bifurcation_condition_1} determines the radius $R_0$ which is the radius of the stationary solution from which the branch of traveling solutions bifurcates. To visualize this bifurcation, we can plot the branches of traveling solutions and stationary solutions together in a bifurcation diagram. The stationary solutions are parameterized by their total total myosin mass $M$ while the traveling solutions are parameterized by their speed $V$, so the bifurcation diagram will have $M$ and $V$ on its axes. The bifurcation occurs when $V=0$ and $M=M_0$, the critical myosin mass obtained by plugging $R_0$ into \eqref{eq:myosin_radius_relation}. The total myosin mass of traveling solutions is calculated as 
\begin{align}
M(V)&=\int_0^{2\pi}\int_0^{R+\rho(\theta,V)}\Lambda(V)e^{\phi(r,\theta,V)-V r\cos\theta} r\,dr\,d\theta\\
&=m_0 \pi R^2+M_2V^2+O(V^4)\label{eq:M_expansion},
\end{align}
where
\begin{equation}
\begin{split}
M_2=2\pi\zeta \int_0^{2\pi}&\frac{m_0 R^2}{\gamma-2k_e R}\phi_2(R,\theta)\,d\theta\\
&+\int_0^{2\pi}\int_0^R\phi_2(r,\theta)r\,dr\,d\theta,
\end{split}
\end{equation}
and $m_0$ comes from \eqref{eq:stationary_m}. An example bifurcation diagram plotting $M(V)$ when $M_2>0$ is shown in Figure \ref{fig:forward_bifurcation}. If $M_2<0$, the branch of steadily moving cells opens to the left instead of the right. The sign of $M_2$ depends on the parameters $R$, $m_0$, $\zeta$, $\gamma$, and $k_e$. Figure \ref{fig:second_myosin_coefficient} shows a typical example of how $M_2$ depends on the effective bulk elasticity $k_e$. In particular, we note that for any combination of parameter values $R$, $m_0$, $\zeta$, and $\gamma$, there are three special values of $k_e$. The first and smallest of these three is $k_0$ at which $M_2=0$. Assuming the higher order terms in \eqref{eq:M_expansion} are bounded in a neighborhood of $k_0$, there exists a small velocity $V_0\neq 0$ such that $M'(V_0)=0$ when $k_e$ is close to $k_0$, giving rise to a \emph{bending point} $(M(V_0),V_0)$ on the branch of traveling solutions, at which $M(V)$ changes from increasing to decreasing (or vice-versa). The second special value of $k_e$ is $k_\ast$ at which $M_2$ is singular, indicating that the bifurcation is not smooth (not twice differentiable) at that parameter value. Finally, the third and largest special value of $k_e$ is $k_c$ which is the smallest value of $k_e$ such that the condition \eqref{eq:monotinicity_condition} holds, which is a requirement for the stability results below.

\begin{figure}
	\centering
	\includegraphics[width=3.4in]{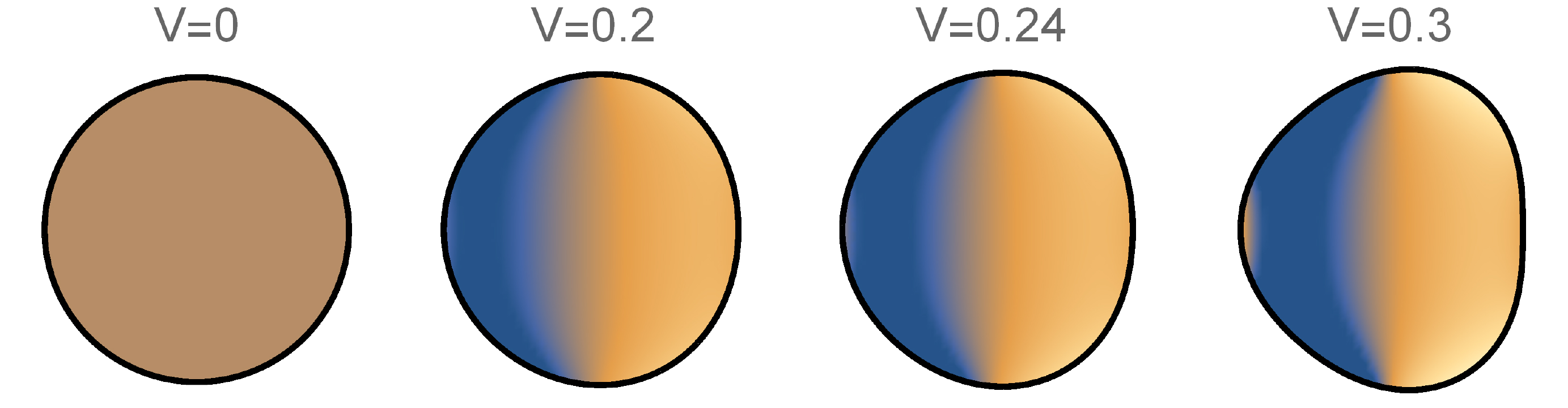}
	\caption{The cell shape and myosin density as speed increases from $V=0$ to $V=0.3$. Motion is to the right. Blue colors indicate higher myosin density.}
	\label{fig:cell_shape}
\end{figure}

In addition to expansion coefficients for scalar stress $\phi_n$, we can solve for the expansion coefficients $\rho_n$ for the bounding curve $R+\rho$ of the region $\Omega$ modeling a cell traveling with constant velocity. We find that $\rho_1=0$, while we find $\rho_2(\theta)=a+b\cos(2\theta)$ and $\rho_3(\theta)=c\cos(3\theta)$ for $a$, $b$, and $c$ depending on physical parameters and the numerically calculated $\phi_2$ and $\phi_3$. Figure \ref{fig:cell_shape} shows the cell shape for various velocities, along with the myosin density $m$ inside the cell. The cell shape and myosin density agree qualitatively with simulations from a similar free-boundary model in \cite{Ker_etal2008} (see Fig. 1) and a phase-field model in \cite{ZieAra2015} (See Fig. 3).

\section{Stability of Traveling Solutions}

In order to have observable steady motion with constant velocity, we need to establish stability of the corresponding traveling solutions. However, the notion of stability of traveling solutions is different from the usual notion of Lyapunov stability. Namely, the best stability we can hope for is \emph{stability up to shifts and change in velocity}, which can be understood as follows. Let $u_0(\mathbf V)$ and $\tilde u_0(\mathbf V)$ be traveling solutions with the same velocity such that $u_0(\mathbf V)$ is centered at the origin and $\tilde u_0(\mathbf V)$ is centered at $(\varepsilon,0)$ for some small $\varepsilon$, i.e., $\tilde u_0(\mathbf V)$ is a \emph{shift} of $u_0(V)$ in the $x$-direction. Since $u_0(\mathbf V)$ and $\tilde u_0(\mathbf V)$ are close, a small perturbation of $u_0(\mathbf V)$ may become close to $\tilde u_0(\mathbf V)$ after a long time. Therefore, \emph{stability up to shifts} means that a perturbation of a traveling solution with velocity $\mathbf V$ stabilizes to another traveling solution with velocity $\mathbf V$, but shifted in the $x$ or $y$ direction. Now consider a traveling solution $u_0(\mathbf V')$ centered at the origin with velocity $\mathbf V'-\mathbf V=\delta$, $\Vert\delta\Vert\ll 1$ at time $t=0$. Then $u_0(\mathbf V)$ and $u_0(\mathbf V')$ have close shapes at $t=0$. After a long time $T\gg 1/\delta$, these solutions are a distance $T\delta\gg 1$ apart. Therefore, a perturbation which is close at $t=0$ to both solutions can only be close to one of the two at $t=T$. This illustrates the concept of \emph{stability up to change in velocity}. Another way of understanding stability up to change in velocity is to observe that this concept is equivalent to stability up to two scalar quantities: \emph{rotation angle} and \emph{speed} (e.g., stability up to rotations is analogous to stability up to shifts since both are coordinate changes). Note that speed is uniquely determined by the total myosin mass $M$, and vice-versa. Therefore, conservation of $M$ can be used to control the speed. We can summarize stability up to shifts and change in velocity in the following way: a perturbation of the traveling solution $u_0(\mathbf V)$ eventually stabilizes to another traveling solution $\tilde u_0(\mathbf V')$ which is initially close to $u_0(\mathbf V)$.

To establish stability up to shifts and change in velocity, we first write \eqref{eq:phi_PDE_1}-\eqref{eq:kinematic_BC_1} as an evolution equation for the myosin density $m$ and polar cell boundary curve $R+\rho$:
\begin{equation}
\frac{\partial}{\partial t}(m,\rho)=F(m,\rho).
\end{equation}
Here $F$ is a nonlinear operator derived from \eqref{eq:m_PDE_1} and \eqref{eq:kinematic_BC_1} (taking $\phi$ as an auxiliary function determined by \eqref{eq:phi_PDE_1} and \eqref{eq:phi_BC_1}). Next, we find the linear operators $A_S(R)$ and $A_T(\mathbf V)$ which are the linearizations of $F$ about the stationary solution at $\mathbf V=0$ with radius $R$ and the traveling solution with velocity $\mathbf V\neq 0$, respectively (to linearize about a constant-in-time solution, $A_T(\mathbf V)$ is found in coordinates moving with velocity $\mathbf V$ so the traveling solution appears stationary). At the bifurcation point, the family of traveling solutions intersects with the family of stationary solutions, so the operators are equal: $A_S(R_0)=A_T(0)$. The signs of the real parts of the eigenvalues of $A_S(R)$ and $A_T(\mathbf V)$ determine the stability of the stationary and traveling solutions. For $R\neq R_0$, all the eigenvalues of $A_S(R)$ are negative except the zero eigenvalue (multiplicity $3$) and an eigenvalue $E(R)$ whose sign depends on $R$, see \eqref{eq:E(R)_formula}. Similarly, for $\mathbf V\neq 0$, all eigenvalues of $A_T(\mathbf V)$ are negative except the zero eigenvalue (multiplicity $4$), and an eigenvalue $\lambda(V)$ whose sign depends on $V=|\mathbf V|$, and determines stability.

Away from the bifurcation point, each of the eigenvectors of $A_S(R)$ (or $A_T(\mathbf V)$) are a derivative of the stationary (or traveling) solution in a parameter (e.g., $R$) with respect to which the class of stationary (or traveling) solutions is invariant. That is, changing this parameter in a stationary (or traveling) solution still results in a stationary (or traveling) solution.
For $R\neq R_0$ and $\mathbf V\neq 0$, both $A_S(R)$ and $A_T(\mathbf V)$ have two eigenvectors for the zero eigenvalue corresponding to invariance with respect to translations in the $x$ and $y$ directions. Additionally, $A_S(R)$ has another eigenvector corresponding to invariance with respect to a change in the radius $R$. For $A_T(\mathbf V)$, the other eigenvectors for the zero eigenvalue are \emph{generalized} eigenvectors because their corresponding parameters change the velocity $\mathbf V$ of a traveling solution--a traveling solution with velocity  $\mathbf V'\neq\mathbf V$ is not constant-in-time in the coordinates that move with velocity $\mathbf V$. These two generalized eigenvectors correspond to changing the speed $V=|\mathbf V|$ of traveling solutions and changing the direction of motion. 

Since the eigenvectors of $A_T(\mathbf V)$ corresponding to shifts both have eigenvalue zero, any linear combination of these two is also an eigenvector. Denote by $a(\mathbf V)$ such a linear combination that has unit length and corresponds to shifts in the direction of $\mathbf V$. Denote by $b(\mathbf V)$ the generalized eigenvector corresponding to invariance in speed $V$ so that $A_T(\mathbf V)b(\mathbf V)=a(\mathbf V)$. These two eigenvectors are of  particular interest because of their relationship with $\lambda(V)$, the  eigenvalue of $A_T(\mathbf V)$ which is the deciding factor in the stability of traveling solutions since it is the one eigenvalue which may be positive. (By the rotational symmetry of $A_T(\mathbf V)$, its eigenvalues depend only on speed $V$, not the direction of $\mathbf V$.) Specifically, we will see that the eigenvector $c(\mathbf V)$ for $\lambda(V)$ becomes parallel to $a(\mathbf V)$ as $V\to 0$. This fact emphasizes the non-self-adjoint nature of $A_T(V)$. Indeed, in the usual self-adjoint case, eigenvectors are orthogonal to one another, so the fact that $a(\mathbf V)$ and $c(\mathbf V)$ are asymptotically parallel is surprising. Since $a(0)$ is parallel to $c(0)$, we conclude that their eigenvalues are also equal: $\lambda(0)=0$. Next, we expect $\lambda(V)$ to depend only on the cell's speed, not its direction, so $\lambda(V)=\lambda(-V)$ and thus $\lambda'(0)=0$. Therefore, we have the asymptotic expansion $\lambda(V)=\lambda_2 V^2+O(V^4)$. Thus, for small $V$, the sign of $\lambda(V)$ is the same as the sign of $\lambda_2$, and the question of stability hinges on this coefficient.

Both $a(\mathbf V)$ and $b(\mathbf V)$ are known explicitly in terms of traveling solution approximated in the previous section, but $\lambda_2$ is not, and neither is $c(\mathbf V)$. Therefore, to find $\lambda_2$, we use an ansatz for $c(\mathbf V)$ derived from its relation to $a(\mathbf V)$ and $b(\mathbf V)$:
\begin{equation}\label{eq:ansatz_for_c}
c(\mathbf V)=a(\mathbf V)+\lambda_2 V^2 b(\mathbf V)+O(V^3).
\end{equation} 
Plugging \eqref{eq:ansatz_for_c} into $A_T(\mathbf V)c(\mathbf V)=\lambda(\mathbf V)c(\mathbf V)$ and comparing terms of like power in $V$, we can solve for $\lambda_2$, though doing so requires solving equations up to fifth order in $V$. We discover that $\lambda_2$ bears an interesting and enlightening relationship with the eigenvalue $E(R)$ of $A_S(R)$ and the total myosin mass $M(V)$ of traveling solutions with velocity $V$:
\begin{equation}\label{eq:lambda_2}
\lambda_2=-\frac{dE}{dR}\Big|_{R=R_0}\frac{dR}{dM}\Big|_{M=M(0)}\frac{d^2 M}{dV^2}\Big|_{V=0}.
\end{equation}

Therefore, the question of stability comes down to the signs of the three derivatives in \eqref{eq:lambda_2}. First $dR/dM$ can be calculated explicitly from \eqref{eq:myosin_radius_relation} and is always negative when \eqref{eq:monotinicity_condition} is met. Next, we observe $\partial^2 M/\partial V^2|_{V=0}=2M_2$, so its sign can be seen in Figure \ref{fig:second_myosin_coefficient}. In particular, $\partial^2 M/\partial V^2|_{V=0}>0$ when condition \eqref{eq:monotinicity_condition} is met. Finally, $dE/dR$ is also explicitly known:
\begin{equation}\label{eq:E(R)_formula}
\frac{dE}{dR}=-C \frac{\partial F}{\partial R},\;\;E(R_0)=0
\end{equation}
where $C>0$ depends on $R$ and the physical constants, and $F$ is given explicitly by \eqref{eq:bifurcation_condition_1}. Observe that $E(R)$ is the largest nonzero eigenvalue for the stationary solutions and it describes their \emph{moveability} in the following sense. If $\text{Re}\,E(R)>0$, then stationary solutions ``want to move.'' Otherwise, they do not. Figure \ref{fig:second_myosin_coefficient} shows how $\partial F/\partial R|_{R=R_0}$ depends on $k_e$. We see that $\partial F/\partial R|_{R=R_0}=0$ precisely when $k_e=k_\ast$, when $M_2$ has a singularity. When condition \eqref{eq:monotinicity_condition} is met ($k_e>k_c>k_\ast$), $\partial F/\partial R|_{R=R_0}>0$, and thus, $\partial E/\partial R|_{R=R_0}<0$. Of the three derivatives in \eqref{eq:lambda_2}, two of them are negative, so $\lambda_2<0$.


Since $\lambda_2<0$, we conclude that all eigenvalues of $A_T(\mathbf V)$ have negative real part except the zero eigenvalue. The eigenvectors for the zero eigenvalue correspond to shifts in the $x$ and $y$ direction and changes to a traveling solution's velocity. Therefore, our analysis suggests that traveling solutions are stable up to shifts and change in velocity, as explained above. Although the velocity of traveling solutions to which our perturbed solution asymptotically stabilizes is not known, its speed is completely determined by the total myosin mass $M$, which is conserved in time. In other words, the asymptotic velocity of a perturbed traveling solution is determined up to a rotation.

\textbf{Conclusions.} We proposed a 2D free boundary model for cell motility which has traveling solutions. We performed linear stability analysis of the traveling solutions. Linearizing about the traveling solution of velocity $\mathbf V$ presented us with two challenges: the linearized problem is not self-adjoint and it has zero eigenvalue of multiplicity four with both true and generalized eigenvectors. The four eigenvectors corresponding to the zero eigenvalues correspond to shifts in the $x$ and $y$ direction and to changes in speed and rotation angle of traveling solutions. There is only one eigenvalue, $\lambda(V)$, which may have positive real part. This led us to introducing a notion of stability \emph{up to shifts and rotation angle (translation and rotation of coordinates)}
, since usual Lyapunov stability does not apply. Moreover, the non-self-adjoint nature of the linearized problem manifests in the eigenvector $c(\mathbf V)$ for $\lambda(V)$ becoming asymptotically parallel to one of the shift eigenvectors. This leads to our specific choice of ansatz for $c(\mathbf V)$, employing a shift eigenvector and its corresponding generalized eigenvector as coefficients in an asymptotic expansion. This ansatz yields an explicit asymptotic formula for $\lambda(V)$ in terms of physical quantities which can be calculated numerically. We found numerically that for most physical parameters, $\text{Re}\,\lambda(V)<0$ for small $V$. Since one can show that all other nonzero eigenvalues have negative real part, and the eigenvectors corresponding to zero correspond to shifts, rotations, and changes in speed, we establish stability of traveling solutions up to translations and rotations of the coordinate system.


\textbf{Acknowledgemennts.} The work of L. Berlyand and C. A. Safsten was partially supported by NSF grant DMS-2005262. V. Rybalko is grateful to PSU Center for Mathematics of Living and Mimetic Matter, and  to PSU Center for Interdisciplinary Mathematics  for support of his  two stays at Penn State. His travel was also supported by NSF grant DMS-1405769. We thank our colleagues I. Aronson, J. Casademunt, J.-F. Joanny, N. Meunier, A. Mogilner, J. Prost, R. Alert, and L. Truskinovsky for many useful discussions on our results and suggestions on the model. We also express our gratitude to the members of the L. Berlyand's PSU research team R. Creese and M. Potomkin for discussions.

\end{document}